# Turbulent Model of Trace Gas Flux in Boundary Layer

Vasenev I.I.*, Nurgaliev I.S.**

*)Head of Ecology Chair, Head of Laboratory Agro-ecological Monitoring, Modeling and Forecasting in Ecosystems, Russian State Agrarian University – Moscow Agricultural Academy named after K.A.Timiryazev

**)Leading Researcher at the same Laboratory

Address: 49, Timiryazevskaya Str., Moscow, 127550, Russian Federation

E-mail: ildus58@mail.ru

## Abstract

Mathematical model of the turbulent flux in the three-layer boundary system is presented. Turbulence is described as a presence of the nonzero vorticity. Generalized advection-diffusion-reaction equation is derived for arbitrary number components in the flux. The fluxes in the layers are objects for matching requirements on the boundaries between the layers.

## Scientific problem

Modeling carbon dioxide and other green house gases (GHG) budgets at the local, regional and global scales is fundamental task in understanding the current and future behavior of the climate cycles and in working out strategies in decision making for sustainable development [1]. For the current tactic purposes the modeling provides potential and instruments for verification emission reduction claims at different levels. The challenges in realistic modeling always were large, especially in the presence of such complex phenomenon as turbulence [2]. Further complications have to do with the variability of fluxes in mixed landscapes, physical aspects in measuring covariance of fluxes and complex atmospheric layer dynamics and the resulting problems of interpretation, representativeness gap filling, footprint and aggregation of data and others. Therefore, it is imperative to clarify and to account all of the key constraints provided/imposed by different data streams, and by processing of ecosystem behavior (natural and manmade) monitoring under the conditions of the complex turbulent atmospheric dynamics as embodied in state-of-the-art models.

**Task formulation**

As usual, we represent system as simple as possible but not simpler. Atmosphere, canopy and soil are represented locally by flat horizontal layers (picture 1). Dominated transport mechanisms identified on the Picture 1.

The substances fluxes within the layers have to be connected between each other by matching conditions as it is required in the well developed theory of partial differential equations. Fluxes' modeling encounters series of frontier complexities, one of which is incomplete knowledge on the nature of turbulence. Eddy covariance method serves as

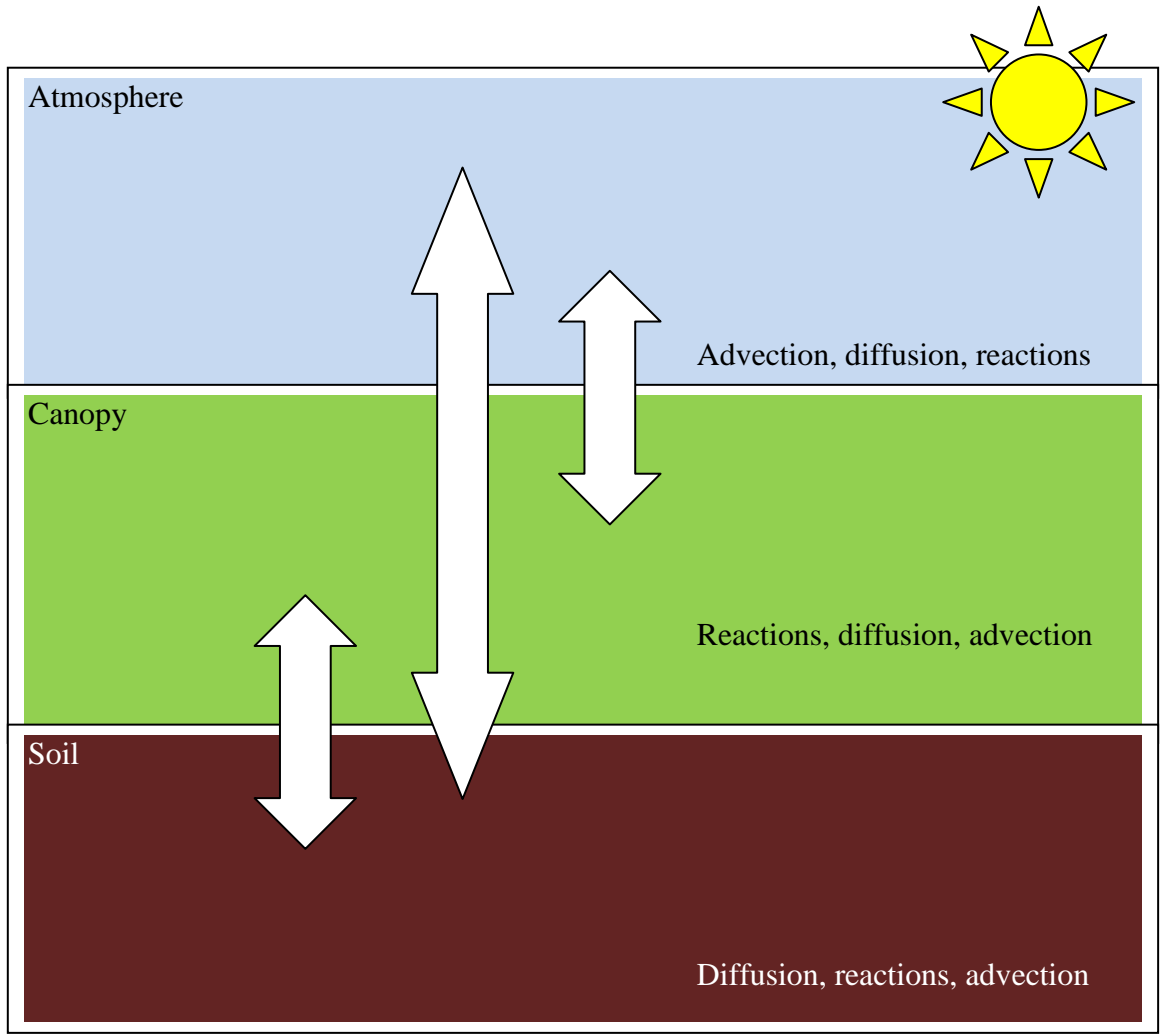

Picture 1. The components of the model.

an available good temporary compromise answer to multiple open questions in the turbulent flux measurement and monitoring.

**GHG flux model equations**

Typical formulae used in calculating eddy flux variables are following [3]

$$u_* = \overline{w'U'} = \frac{1}{N}\sum_{i=1}^{N} w'U'$$

$$\overline{w'T'} = \frac{1}{N}\sum_{i=1}^{N} w'T'$$

$$\overline{w'x'} = \frac{1}{N}\sum_{i=1}^{N}\left(\overline{w'x'} + \frac{m_d}{m_w}\frac{\rho_x}{\rho_d}\overline{w'\rho'_w} + \rho_x\left(1+\frac{m_d}{m_w}\frac{\rho_x}{\rho_d}\right)\frac{\overline{w'T'}}{T}\right)$$

where $U$ is the horizontal wind speed, $w$ is the vertical wind speed, $T$ is the air temperature, $x$ is any scalar (e.g. the the $CO_2$ or any other GHG component concentration, or - specific humidity $q$, etc.), $m_d/m_w$ is the ratio of molecular weight of the dry air to that of water vapor, and $\rho_d$, $\rho_w$, and $\rho_x$ are the densities for dry air, water vapor, and $x$ respectively. The right hand side of $\overline{w'x'}$ has extra terms due to corrections in air density [4,5]. See also other corrections [6].

Key phenomenon in the lower layer of atmosphere where the GHG flux is formed and measured is turbulence. New approach to flux modeling in this article is based on modification of the material particle concept. We will call it material point of the second type meaning that traditional concept of the material point as a small body with the negligible size but with a given amount of mass will be called material point of the first type.

## New Concept of Material Point

In the traditional introductions into eddy covariance method one of basic primary concepts is an idea of *__fluid parcel__* which "is a very small amount of fluid, identifiable throughout its dynamic history while moving with the fluid flow" (wikipedia). Other close ideas are material point, particle. We introduce other conception called material point of the second type. The difference of the new conception from the traditional conception of the material point (let us call it material point of the first type) is following. The point of the second type is not characterized by its mass but density, its motion is a motion of the continuous media – with deformations and torsion. Hereby we have said goodbye to the old sophism "let us call material particle material point and will treat it as a point." Sorry, material point is material. Therefore it has inalienable ability to be oriented relative to other material objects. And calling it point should not eliminate this essential ability. Doing so is sophism.

This revelation brings as to reconsideration of the flux. A new aspect of the new model is awareness on necessity of projecting velocity vector $V_\alpha$ to configuration space of radius-vectors $R^\beta$ to manipulate with velocities Cartesian coordinates:

$$V_\alpha = H_{\alpha\beta} R^\beta, \quad (1)$$

because velocity vector is an object of the different (tangential) space. For the fundamental geometrical aspects of this statement see [7].

Here $H_{\alpha\beta}$ is advection-distortion-vortex tensor (affinor), $R^\beta$ is radius-vector of the material point. Splitting tensor $H_{\alpha\beta}$ into three parts, responsible for expansion $\theta\delta_{\alpha\beta}/3$ shear $\sigma_{\alpha\beta}$ and vorticity $\omega_{\alpha\beta}$ is a novel tool to describe turbulence in the boundary layer agro-meteorology accounting, new degrees of freedom of the particle modeled as a material point of the second type:

$$H_{\alpha\beta} = \theta\delta_{\alpha\beta}/3 + \sigma_{\alpha\beta} - \omega_{\alpha\beta}. \quad (2)$$

**GHG flux model equations**

Linear hydrodynamic Euler equations split into

$$\frac{d\theta}{dt} + \frac{1}{3}\theta^2 + \sigma^2 - \omega^2 = -F_{\alpha\alpha}$$

$$\sigma^2 = \sigma_{\alpha\beta}\sigma_{\alpha\beta}, \omega^2 = \omega_{\alpha\beta}\omega_{\alpha\beta}, \frac{d\rho}{dt} = -\rho\vartheta,$$
(3)

$$\frac{d\sigma_{\alpha\beta}}{dt} + \frac{2}{3}\vartheta\sigma_{\alpha\beta} + \sigma_{\alpha\gamma}\sigma_{\gamma\beta} - \frac{1}{3}\delta_{\alpha\delta}\sigma^2 + \omega_{\alpha\gamma}\omega_{\gamma\beta} + \frac{1}{3}\delta_{\alpha\delta}\omega^2 = -F_{(\alpha\beta)} + \frac{1}{3}\delta_{\alpha\delta}F_{\gamma\gamma}$$

$$\frac{d\omega_{\alpha\beta}}{dt} + \frac{2}{3}\vartheta\omega_{\alpha\beta} + \sigma_{\alpha\gamma}\omega_{\gamma\beta} + \sigma_{\alpha\gamma}\omega_{\gamma\beta} + \omega_{\alpha\gamma}\delta_{\gamma\beta} = -F_{[\alpha\beta]}.$$

Here $F_{\alpha\beta}$ is a tensor of the external forces gradient. Dynamics of the turbulent flux (3) is much more complex than of laminar flux and flux measurement methods need including turbulence-related terms.

### Advection-diffusion-reaction equations

For a more general description structure froming (e.g. eddies) formation we get equation of the multi-component reaction-diffusion type

$$\frac{\partial x_i}{\partial t} = f_i(\{x_j\}) + \nabla_k D^k{}_j(r)\nabla^j x_i.$$

Substances diffusion and transfer of thermal energy are described by the same class equations. This equation is extremely universal and can be applied in modeling the broad range of processes taking place in agroindustry and in its energetics. The most impressive new application is a thermochemical decomposition of organic material such us agro manure at elevated temperatures without the participation of oxygen (pyrolysis) and production of designed fuels. It goes without saying that migration and generating (reacting) of traces gases in soil described by this equation.

The geometries of the subsystems are also considered in the modeling of the growth kinetics as a crucial factor. New class of equations called advection-diffusion-reaction (http://arxiv.org/pdf/1210.4091v2.pdf) was derived as following

$$\frac{\partial \rho_i}{\partial t} = -\tfrac{1}{3} H_i \rho_i + f(\{\rho_i\}) + \nabla\left[D_i(\{\rho_i\})\nabla \rho_i\right].$$

This equation describes arbitrary amount of material components with densities $\rho_i$, parameters $H_i$ - diagonal elements of matrix in (1) - are responsible for an advective change of density, and coefficients of the effective diffusion $D_i$, generalized and adopted when needed. It may also provide nonlinear evolution scenarios for evolution of the multi-component reacting media in the different systems such as agro-bio-geo fluxes. The nonlinear term $f(\{\rho_i\})$ stands for reactions between the components. The fluxes in the layers are objects for matching requirements on the boundaries between the layers. These requirements are fulfilled by appropriate identification of the constants of integration.

One of the new features of the nonlinear dynamic processes described by given equation is the existence of the so called threshold effects. This means that we may expect emergence and ability to long existence of some eddies and grow some of them to scales intensities of tornado before getting destroyed up.

Next aspect in using physical laws in the presence of turbulence is taking into account vorticity in Doppler Effect for measurement of speed of flux. The basic equation describing (ultra)sound in turbulent flux is Blokhintsev- Howe equation [8,9]. In the adiabatic approximation the equation holds

$$\left\{\frac{D}{Dt}\left(\frac{1}{c^2}\frac{D}{Dt}\right) + \frac{1}{c^2}\frac{Dv}{Dt}\nabla - \nabla^2\right\}B = \text{div } L - \frac{1}{c^2}\frac{Dv}{Dt}L,$$

$$B = H + v^2/2, \quad \frac{D}{Dt} = \frac{\partial}{\partial t} + v\nabla, \quad L = \Omega \times v - T\nabla S.$$

Here $H$ is enthalpy, $\Omega = \mathbf{rot}v$ is vorticity, $S$ - entropy, $T$ - temperature, $v$ is flux speed, $c$ is sound speed, $t$ is time. Note that drag enthalpy $B$ is connected to sound pressure $p$ as $\partial p/\partial t = \rho DB/Dt$ ($\rho$ is mass density of the media). Blokhintsev- Howe equation is derived as a consequence of impulse and mas concervation as well as equation of the state of the ideal gas. LHS of it correspond to the transfer of the sound in the arbitrary non-homogeneous flux and RHS chracterizes the sources of the sound connected to character of the flux such as presence of vorticity and entropy gradient. As one can see, sound equation much more complicated than those traditionally used for deriving Doppler Effect in the frame of eddy covariance method.

## Conclusions

Dynamics of the turbulent GHG flux on the boundary layer, as given mathematical model demonstrates, is much complicated in comparison with vortex-free motion. Therefore, eddy covariance method, as an existing broadly accepted successfully working instrument, needs further development to bring even more detailed models, accounting vorticity. Vorticity is the phenomenon changing radically many of the scenarios in nature [9,10]. Therefore we may expect that existing questions, not last of which is energy balance problem, can be tackled using vortex.

## Acknowledgements

This research was funded by Government of Russian Federation, Grant № 11.G34.31.0079.